\documentclass[authoryear, 3p]{elsarticle}
\usepackage[T1]{fontenc}
\usepackage{rotating}
\usepackage{lipsum}
\usepackage{graphicx}
\usepackage{setspace}
\graphicspath{{/Users/nick/mypapers/C3/images/}}
\usepackage[nomarkers,figuresonly]{endfloat}
\title{Spatial and Seasonal Variations in C$_{3}$H$_{x}$ Hydrocarbon Abundance in Titan's Stratosphere from Cassini CIRS Observations}
\author[gsfc,umbc]{Nicholas A Lombardo\corref{cor1}}
\ead{nicholas.lombardo@nasa.gov}
\author[gsfc]{Conor A Nixon}
\author[gsfc,umcp]{Richard K Achterberg}
\author[lisa]{Antoine Jolly}
\author[jpl]{Keeyoon Sung}
\author[ox]{Patrick G J Irwin}
\author[gsfc]{F Michael Flasar}
\cortext[cor1]{Corresponding author: Nicholas Lombardo, nicholas.lombardo@nasa.gov}
\address[gsfc]{Planetary Systems Laboratory, Solar System Exploration Division, NASA Goddard Space Flight Center, 8800 Greenbelt Road, Greenbelt, MD, USA}
\address[umbc]{Center for Space Science and Technology, University of Maryland, Baltimore County, 1000 Hilltop Circle, Baltimore, MD, USA}
\address[umcp]{Department of Astronomy, University of Maryland College Park, College Park, MD, USA}
\address[lisa]{Laboratoire Interuniversitaire des Syst\'emes Atmosph\'eriques, Universit\'e Paris-Est, Creteil, France}
\address[jpl]{Jet Propulsion Laboratory, California Institute of Technology, Pasadena, CA, USA}
\address[ox]{Atmospheric, Oceanic and Planetary Physics, Clarendon Laboratory, University of Oxford, Parks Road, Oxford OX1 3PU, UK}
\corref{}

\begin{document}
\begin{abstract}
	Of the C$_{3}$H$_{x}$ hydrocarbons, propane (C${_3}$H$_{8}$) and propyne (methylacetylene, CH$_{3}$C$_{2}$H) were first detected in Titan's atmosphere during the Voyager 1 flyby in 1980.  Propene (propylene, C$_{3}$H$_{6}$) was first detected in 2013 with data from the Composite InfraRed Spectrometer (CIRS) instrument on Cassini.  We present the first measured abundance profiles of propene on Titan from radiative transfer modeling, and compare our measurements to predictions derived from several photochemical models.  Near the equator, propene is observed to have a peak abundance of 10 ppbv at a pressure of 0.2 mbar. Several photochemical models predict the amount at this pressure to be in the range 0.3 - 1 ppbv and also show a local minimum near 0.2 mbar which we do not see in our measurements.  We also see that propene follows a different latitudinal trend than the other C$_{3}$ molecules.  While propane and propyne concentrate near the winter pole, transported via a global convective cell, propene is most abundant above the equator.
	We retrieve vertical abundances profiles between 125 km and 375 km for these gases for latitude averages between 60$^{\circ}$S to 20$^{\circ}$S, 20$^{\circ}$S to 20$^{\circ}$N, and 20$^{\circ}$N to 60$^{\circ}$N over two time periods, 2004 through 2009 representing Titan's atmosphere before the 2009 equinox, and 2012 through 2015 representing time after the equinox.  
	
	Additionally, using newly corrected line data, we determined an updated upper limit for allene (propadiene, CH$_{2}$CCH$_{2}$, the isomer of propyne).  We claim a 3-$\sigma$ upper limit mixing ratio of 2.5$\times$10$^{-9}$ within 30$^\circ$ of the equator.
	The measurements we present will  further constrain photochemical models by refining reaction rates and the transport of these gases throughout Titan's atmosphere.

\end{abstract}
\maketitle
	\section{Introduction}	
	Titan, the largest moon of Saturn, has a CH$_{4}$ surface mixing ratio of about 5\%, measured by the Huygens GCMS \citep{niemann:2010}, and decreasing with altitude into the stratosphere where it remains constant with altitude at 1-1.5\%, as measured in \cite{lellouch:2014}.   Titan is thought to have many similarities to the Archean Earth, including an atmosphere abundant in in N$_{2}$ and significant quantities of CH$_{4}$ as well as global haze layers which continually shroud Titan and may have occurred intermittently on Earth.  While factors like temperature, sources of atmospheric CH$_{4}$, and minor atmospheric constituents vary between the two bodies, Titan remains a good analog for studying the atmosphere of the Archean Earth \citep{arney:2016, izon:2017}.
	
	The global haze on Titan is produced through photolysis of CH$_{4}$ as Saturn Magnetospheric Electrons and solar UV photons bombard the upper atmosphere.  The products of this process -highly reactive CH$_{3}^{-}$, H$^{+}$, and N$^{+}$ ions, among others- may then react to form C$_{2}$H$_{6}$, C$_{2}$H$_{4}$, and other molecules.  As this complex process continues,  larger hydrocarbons (C$_{x}$H$_{y}$) and nitriles (C$_{x}$H$_{y}$(CN)$_{z}$) react further to give rise to the 'photochemical zoo' of molecules present in Titan's atmosphere \citep{yung:1984, wilson:2004,  lavvas:2008, loison:2015, dobrijevic:2016, willacy:2016}.
	
	Titan's 26.7$^{\circ}$ obliquity (the axial tilt relative to the normal of the orbital plane), comparable to the Earth's 23.5$^{\circ}$ obliquity, causes variations in the insolation of the moon over the course of a Titan year (about 29.5 Earth years). The resulting seasonal variations in the physical state of the atmosphere include molecule abundance \citep{vinatier:2015, coustenis:2018}, temperature \citep{achterberg:2011, teanby:2017}, and behavior of the haze layers \citep{jennings:2012}, discussed more in the review by \cite{horst:17}.  Noteworthy is the existence of a global circulation cell, which transports warm gases in the summer hemisphere towards the winter pole, where they subside lower into the stratosphere.  This downward advection causes adiabatic warming in the winter stratosphere and entrains short-lived gases produced in the upper stratosphere, increasing their abundance lower in the atmosphere.  As northern winter evolved to northern spring, this single circulation cell transformed into two circulation cells, upwelling near the equator and downwelling at both poles, as predicted in \cite{hourdin:2004} and observed in \cite{teanby:2012}.  For additional explanation of Titan's atmospheric dynamics and chemistry, the reader is directed to \cite{titan:2010} and \cite{titan:2014}.
	
	Regarding the C$_{3}$ hydrocarbons, propane (C$_{3}$H$_{8}$) and propyne (C$_{3}$H$_{4}$) were initially detected in Titan's atmosphere after the 1980 Voyager 1 flyby  \citep{hanel:1981} through spectra acquired by the IRIS instrument.  Abundances for propyne were first estimated by \citep{maguire:1981} by comparing the strength of the 633 cm$^{-1}$ Q-branch of propyne to the 721 cm$^{-1}$ Q-branch of acetylene (also ethyne, C$_{2}$H$_{2}$), and estimated to be on the order of 3$\times$10$^{-8}$. Propane was modeled in the same paper using a synthetic spectrum constructed for its $\nu_{21}$ band, and a disk averaged value of 2$\times$10$^{-5}$ was reported.  These values were updated by \cite{coustenis:1989} to 4.4$^{+1.7}_{-2.1}\times$10$^{-9}$ for propyne and (7$\pm$4)$\times$10$^{-7}$ for propane.  Further weak bands of propane were detected by the Composite InfraRed Spectrometer (CIRS) aboard Cassini \citep{nixon:propane}.  Over three decades later, CIRS spectra were used to make the first detection of C$_{3}$H$_{6}$  \citep{nixon:propene}, however an exact abundance could not be retrieved from modeling the spectra due to the lack of a spectral line list, although an abundance estimate was made by comparing the intensities of propene and propane lines, discussed more in Section 4.2.  
	
	Recent analyses have shown the abundance of propyne to vary strongly with season and latitude.  \cite{vinatier:2015}, using limb viewing observations, showed the vertical gradient of C$_{3}$H$_{4}$ increases dramatically over the mid northern latitudes as northern winter moves into northern spring and the polar vortex responds to the changing amount of sunlight. 
	  \cite{coustenis:2018}, using nadir observations to probe abundance in a narrow altitude range in the middle stratosphere, show a similar trend at latitudes closer to the pole, between 60$^{\circ}$ and 90$^{\circ}$ either side of the equator.  In the same studies, propane was shown to have a more constant abundance in latitude and time, remaining constant within error bars near 1$\times$10$^{-6}$ throughout the stratosphere, with the exception near the winter pole, where it increases with altitude.
		
	Two C$_{3}$ hydrocarbons have yet to be firmly detected on Titan, allene (CH$_{2}$CCH$_{2}$, isomer of propyne) and cyclopropane (CH$_{2}$CH$_{2}$CH$_{2}$, isomer of propene). There was a tentative detection of allene by \cite{roe:2011}, however an accurate line list was not available at the time of the study, thus the authors were not able to model the potential allene feature and confirm its detection.  In this paper, we discuss members of the C$_{3}$H$_{x}$ series known to be present in Titan's atmosphere- propane (C$_{3}$H$_{8}$), propene (C$_{3}$H$_{6}$), and propyne (CH$_{3}$C$_{2}$H).  We also searched for allene and provide an new upper limit for allene in regions close to the equator.  This work was enabled by the creation of a propene pseudo-line list for Titan \citep{sung:18} and an updated line list for allene (see Section 2.2).                                                                                                                                                                                                                                                                                                                                                                                                                                                                                                                                                                                                                                                                                                                                                                                                                                                                                                                                                                                                                                                                                                                                                                                                                                                                                                                                                                           
	
	We use spectra collected by the CIRS instrument to determine the abundance of propene in Titan's stratosphere. We show latitudinal and seasonal variation in the distribution of propene, propane and propyne. The large number of CIRS observations used allows us to vertically resolve the profile of each gas. We compare the values determined to those predicted by photochemical models of \cite{hebrard:2013}, \cite{kras:2014}, \cite{li:2015}, and \cite{loison:2015}. Additionally, we use a corrected line list for allene to determine an updated upper limit for the molecule.
	
	\section{Methods}
	\subsection{Dataset} 
	CIRS is a Fourier Transform infrared spectrometer, with three focal planes spanning the 10 cm$^{-1}$ - 1500 cm$^{-1}$ spectral region \citep{jennings:17}.  We use spectra acquired by Focal Plane 3 (FP3, 580 cm$^{-1}$-1100 cm$^{-1}$) and Focal Plane 4 (FP4, 1050 cm$^{-1}$-1500cm$^{-1}$ ), two parallel arrays of 10 detectors each.  Limb observations were performed at a spectral resolution of 0.5 cm$^{-1}$ at distances between 10$^{5}$ km and 2$\times$10$^{5}$ km from Titan, during which time each focal plane was positioned normal to Titan's surface, such that each detector sampled a different altitude.  The arrays were centered at ~125 km for between one and two hours and were then moved away from Titan's surface to stare at a central altitude of ~350 km for a similar amount of time.  The footprint of each detector (the vertical resolution) on Titan's atmosphere varied between 27 km and 54 km depending on the distance to the moon.  The size of the footprint was comparable to Titan's atmospheric scale height, and thus allows us to vertically resolve physical characteristics of the atmosphere.
	
	The C$_{3}$H$_{6}$ $\nu_{19}$ band detected in \cite{nixon:propene} at 912.67 cm$^{-1}$ sits between several C$_{2}$H$_{4}$ emissions, and on top of a broad C$_{3}$H$_{8}$ band.  To increase S/N to the point where we can model this feature, we divide the CIRS dataset into six time-latitude bins, where in each bin the temperature varies less than 15 K.  We make the assumption that over the time and latitudes covered in by each bin, Titan's stratosphere has similar temperature and molecular abundance profiles. Spectra in each bin are averaged together to reduce random noise in the data before modeling.  Each bin includes data from from between three and seven flybys (or between 187 and 728 spectra).  We use two time periods - the pre equinox time from 2005 to 2009 (just before the northern vernal equinox of August 2009) and the post equinox time 2012 through 2015.  During the time just after the northern vernal equinox, Cassini was in a very low inclination orbit relative to Saturn.  Limb observations on Titan were focused on the polar regions, as Cassini was able to view these regions of the atmosphere continuously during a Titan flyby.  We therefore do not include data from just after the equinox, as no limb data exists for the latitude regions we model in this work.  In each time span, we combine observations representative of three latitude regions - northern, equatorial, and southern.  
	
\begin{figure}[p]
	\includegraphics[width=\columnwidth]{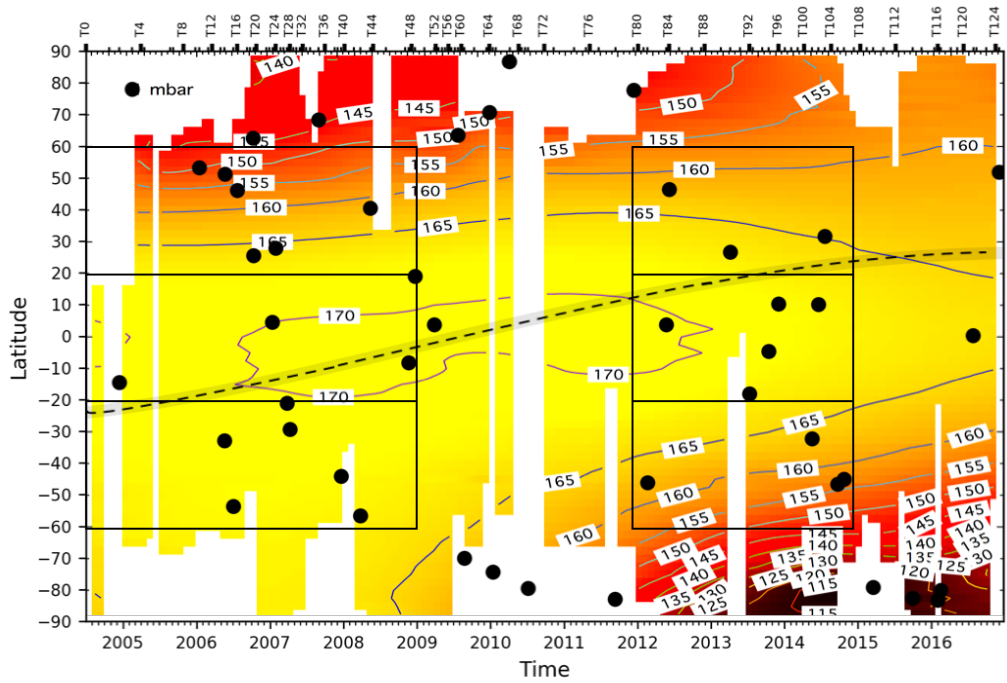}
	\caption{Distribution of all Mid-IR Limb Integrations (MIRLMBINT) observations during the Cassini mission shown as black circles.  Temperatures are shown as contours (colored) and correspond to those at 1 mbar, or about 175 km, and are updated from \cite{achterberg:2014}.  The black boxes indicate the binning scheme with two time spans (2004-2009 and 2012-2015) further binned to three latitude ranges (60$^{\circ}$S-20$^{\circ}$S, 20$^{\circ}$S-20$^{\circ}$N, and 20$^{\circ}$N-60$^{\circ}$N).  The dashed black line surrounded by gray is the solar latitude of Titan, where the Sun is directly overhead at 1200 local time.}
	\label{fig:map}
	\centering
	\end{figure}

	In our averages, we include data from 20$^{\circ}$ to 60$^{\circ}$ for both hemispheres, and within 20$^{\circ}$ of the equator.  The former two bins represent the mid-latitudes, and the third bin represent the equatorial atmosphere.  While the latitude boundaries for observations we include are the same before and after equinox, the physical distribution of included observations varies.  As an example, in the pre-equinox time, the northern bin contains observations from 24$^{\circ}$N through 54$^{\circ}$N, whereas post-equinox we have observations only between 25$^{\circ}$N and 48$^{\circ}$N.  The distribution of observations used is shown in Fig. \ref{fig:map} as black dots.  The boundaries for each time-latitudinal bin are drawn as black boxes enclosing observations.  We exclude observation centered at latitudes closer than 60$^{\circ}$ to either pole because temperature begins to vary strongly with latitude in these regions.  Including these spectra in our averages would make the resultant spectra very difficult to model and obscure details of finer latitudinal variations in the retrieved profiles.
	
	A summary of the data used is in Table \ref{tab:data}.
	
	\begin{sidewaystable}[p]
		\centering  
		\begin{tabular}{ccccccc|cccccc}
			
			Time Range &\multicolumn{6}{c|}{2004-2009} &\multicolumn{6}{c}{2012-2015}\\
			
			Latitude Range &\multicolumn{2}{c}{60$^{\circ}$S-20$^{\circ}$S} & \multicolumn{2}{c}{20$^{\circ}$S-20$^{\circ}$N} & \multicolumn{2}{c|}{20$^{\circ}$N-60$^{\circ}$N}
			& \multicolumn{2}{c}{60$^{\circ}$S-20$^{\circ}$S} & \multicolumn{2}{c}{20$^{\circ}$S-20$^{\circ}$N} & \multicolumn{2}{c}{20$^{\circ}$N-60$^{\circ}$N}\\
			
			Avg Lat&\multicolumn{2}{c}{42$^{\circ}$S} & \multicolumn{2}{c}{0$^{\circ}$N} & \multicolumn{2}{c|}{43$^{\circ}$N}
			& \multicolumn{2}{c}{60$^{\circ}$S-20$^{\circ}$S} & \multicolumn{2}{c}{1$^{\circ}$N} & \multicolumn{2}{c}{34$^{\circ}$N}\\
			\hline \hline	
			
			\multicolumn{13}{c}{FP3}\\
			
			Altitude (km) & spectra & NESR & spectra & NESR & spectra & NESR  & spectra & NESR & spectra & NESR  & spectra & NESR \\
			
			100-150 & 322 & 1.58 & 330 & 1.68 & 412  & 1.53 & 299 & 2.30 & 440 & 1.82 & 288 & 2.41\\
			
			150-200 & 451 & 1.37 & 428 & 1.48 &  537 & 1.37 & 416 & 1.96 & 608 & 1.67 & 395 & 2.19\\
			
			200-250 & 515 & 1.32 & 460 & 1.50 & 568 & 1.30 & 441 & 1.85 & 601 & 1.61 & 464 & 1.89\\
			
			250-300 & 496 & 1.34 & 440 & 1.45 & 634 & 1.23 & 463 & 1.99 & 680 & 1.60 & 446 & 1.74\\
			
			300-350 & 271 & 1.67 & 299 & 1.79 & 294 & 1.60 & 229 & 2.60 & 297 & 2.16 & 341 & 1.92\\
			
			350-375 & 235 & 1.96 & 242 & 1.89 & 258 & 1.59 &  218 & 2.33 & 287 & 2.21 & 191 & 2.70\\
			\hline
			\multicolumn{13}{c}{} \\
			\multicolumn{13}{c}{FP4}\\
			
			Altitude (km) & spectra & NESR & spectra & NESR & spectra & NESR  & spectra & NESR & spectra & NESR  & spectra & NESR \\
			\hline
			
			100-150 & 310 & 0.29 & 369 & 0.23 & 515  & 0.29 & 388 & 0.40 & 345 & 0.30 & 298 & 0.33 \\
			
			150-200 & 381 & 0.22 & 463 & 0.27 &  652 & 0.23& 530 & 0.27 & 488 & 0.30 & 397 & 0.32\\
			
			200-250 & 372 & 0.24 & 573 & 0.24 & 691 & 0.17& 551 & 0.21 & 503 & 0.26 & 464 & 0.29\\
			
			250-300 & 334 &  0.26& 559 & 0.19 & 728 & 0.18 & 583 & 0.20 & 556 & 0.23 & 450 & 0.24\\
			
			300-350 & 187 &0.39 & 450 & 0.22& 345 & 0.28 & 305 & 0.29 & 230 & 0.33 & 336 & 0.28\\
			
			350-400 & 187 &  0.29& 277 & 0.30 & 294 & 0.30 & 265 & 0.32 & 232 & 0.38 & 183 & 0.29\\
			\hline
			
		\end{tabular}
		
		\caption{Observations averaged.  Listed are the middle altitude, number of spectra averaged, and the noise equivalent spectral radiance (NESR) for each altitude bin.  NESRs have units of nW cm$^{-2}$ sr$^{-1}$ cm, and were derived from the standard deviation of the mean for each average, and were measured at 910 cm$^{-1}$ for FP3, and 1280 cm$^{-1}$ for FP4 .}
		\label{tab:data}
	\end{sidewaystable}

	\subsection{Spectral Line Data}
	Molecular line lists used are described in Table \ref{tab:linedata}.
	
	\begin{table}[h]
		\centering
		\begin{tabular}{llll}
			\hline
			Molecule & \textit{a priori} VMR & Source of Line Data\\
			\hline
			C$_{2}$H$_{4}$		&(1.1$\pm$0.2)$\times$10$^{-7}$ & \cite{geisa:2016}\\
			C$_{3}$H$_{8}$		&(1$\pm$0.5)$\times$10$^{-6}$ & \cite{sung:propane} \\
			C$_{3}$H$_{6}$		&(3$\pm$1.5)$\times$10$^{-9}$ & \cite{sung:18} \\
			CH$_{3}$C$_{2}$H & (1$\pm$0.4) $\times$10$^{-8}$ & same as \cite{coustenis:2007} \\
			C$_{4}$H$_{2}$		&(5$\pm$2.5)$\times$10$^{-9}$ & \cite{jolly:2010}\\
			CH$_{2}$CCH$_{2}$ & (3$\pm$1.5$\times$10$^{-9}$) & Discussed in Section 2.2.1\\
			\hline
		\end{tabular}
		\caption{The reference abundances of molecules retrieved in the model are set to be constant above saturation at the values listed here.  Sources of the spectral line data are also listed.}
		\label{tab:linedata}
	\end{table}
	
	\subsubsection{Updated Allene Linedata}
	No allene line lists are currently present in the HITRAN or GEISA databases.  \cite{coustenis:1993} initially investigated the detectability of allene in Titan's atmosphere using spectroscopic parameters by \cite{chazelas:1985} for the $\nu_{10}$ band centered at 845 cm$^{-1}$.  Line intensities were obtained from band intensity measurements by \cite{koga:1979}. They concluded that the non-detection of allene implied an abundance below 5 ppbv.  The same line list was used in \cite{coustenis:2003} and \cite{nixon:allene}, with upper limits discussed in Section 4.4.
	
	As in \cite{jolly:2015}, we notice that the locations of transitions and transition intensities listed in the older line list does not match those in room temperature laboratory spectra.  In this search for allene, we make the necessary corrections to this older line list so the spectral data we use matches experimental laboratory spectra.
	
\begin{figure}[p]
		\includegraphics[height = \textheight]{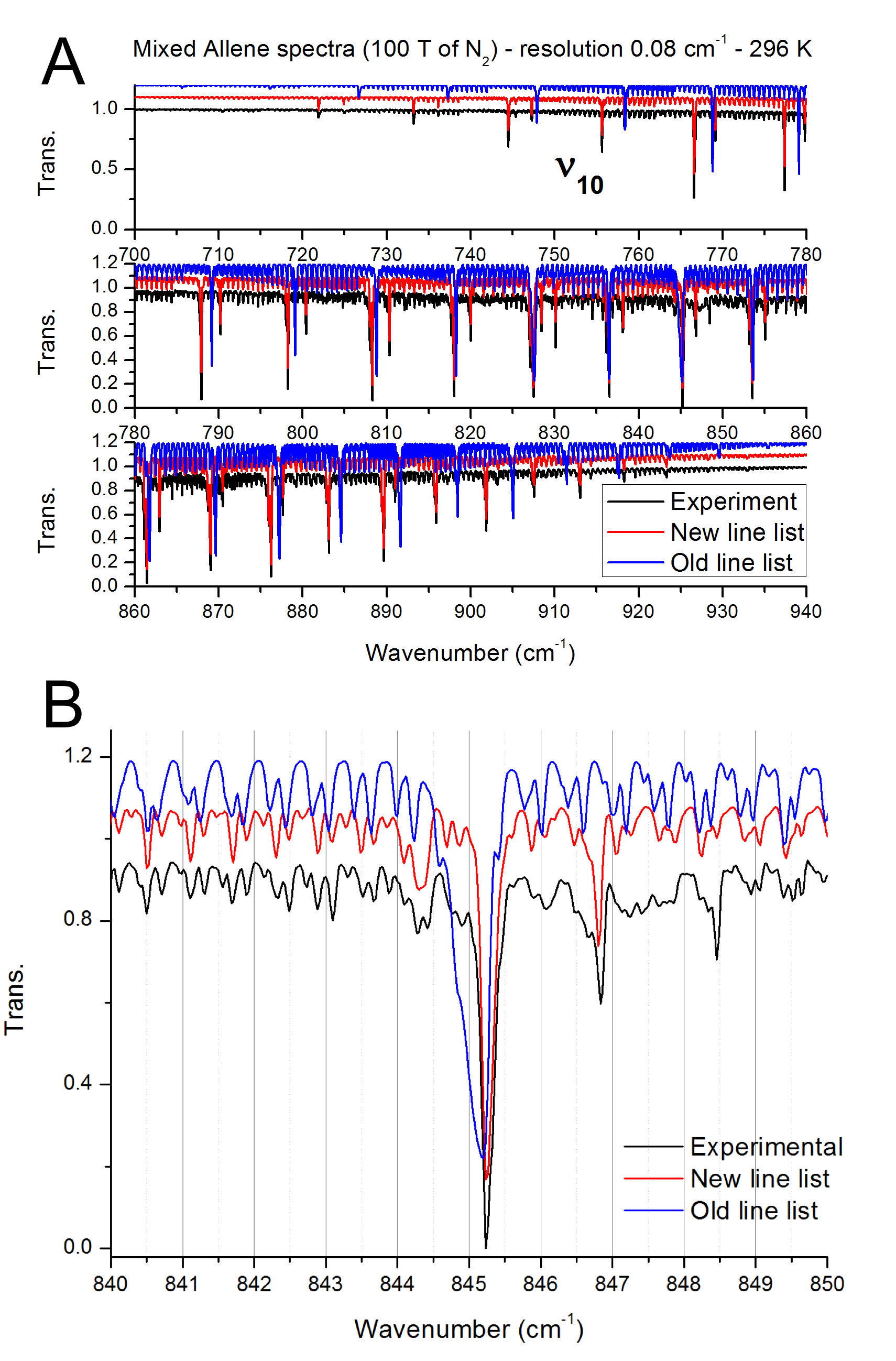}	
		\centering	
		\caption{A: A comparison of the $\nu_{10}$ band of allene in the old line list used in previous works (blue), new line list used in this work (red), and experimental laboratory spectrum (black).  The old line list is not consistent with the laboratory spectrum, and is missing several hot bands.  Many transitions are also at incorrect wavenumbers.  B: a close up of the 845 $cm^{-1}$ region.}
		\label{fig:allene_ll}
	\end{figure}

	In this work, we combine the existing spectroscopic data used previously with parameters obtained by high resolution studies \citep{hegelund:1993} that address the existence of hot bands. The hot band contribution is necessary at room temperature to allow comparisons between these calculated line lists with observed room temperature spectra.  Calculations were compared and validated against room and high temperature spectra taken at 0.08 cm$^{-1}$ resolution in the $\nu_{10}$/$\nu_{9}$ wavenumber range \citep{es-sebbar:2014}.  A comparison between the old line list and new line list are given against a laboratory spectrum in Fig. \ref{fig:allene_ll}, where the misfit between the line list used in previous studies and the laboratory data can be clearly seen.  
	
	\subsection{Radiative Transfer Modeling}

	Spectral fitting was achieved using the NEMESIS atmospheric modeling code \citep{irwin:nemesis}.  NEMESIS operates on the method of optimal estimation, which involves the computation of a forward model and a retrieval process. The forward model was calculated using the correlated-k method of \cite{lacis:1991}, and includes a Hamming apodization of Full Width at Half Maximum of 0.475 cm $^{-1}$ to recreate the instrument line shape.   The retrieval process varies \textit{a priori} profiles of chosen physical parameters to optimize the spectral fits.  This is performed by minimizing a cost-function  which includes the deviation of the retrieved profile from the \textit{a priori} estimate, and the quality of fit to the spectra (similar to a $\chi^{2}$ goodness of fit test).   NEMESIS has been extensively used to determine atmospheric abundances in the outer solar system using IR spectra, and application to Titan is described in \cite{teanby:2007}, \cite{teanby:2009}, and \cite{cottini:2012}.  
	
	Spectral modeling for each bin proceeded in two steps: stratospheric temperatures during observations were extracted and then used in determining the abundances of trace gases.  First, the 1275 cm$^{-1}$ - 1325 cm$^{-1}$ region of the $\nu_{4}$ band of methane was modeled to retrieve the stratospheric temperature profile.  While we did not model the entire $\nu_{4}$ band that extends from 1250 cm$^{-1}$ - 1350 cm$^{-1}$, the region of the spectrum we modeled contains sufficient information to retrieve the temperatures in the altitude regions needed for modeling the trace gases. Temperature measurements proceeded by assuming a methane abundance of 1.41$\times$10$^{-2}$ above 140 km, consistent with previous measurements and models \citep{lellouch:2014},  \citep{wilson:2004}) and an abundance below 140 km derived from measurements by the Huygens descent profile \citep{niemann:2010}. In the retrieval, the temperature profile, derived from the HASI temperature \citep{ful:2005} profile, and a non-gray aerosol haze were allowed to vary to fit the observed spectrum. This retrieved temperature was then fixed while molecules in the FP3 spectral region were allowed to vary. In the 900 cm$^{-1}$ - 930 cm$^{-1}$ region, the Q branch of the $\nu_{21}$ band of propane at 922 cm$^{-1}$, the $\nu_{19}$ propene band at 912.5 cm$^{-1}$, and $\nu_{7}$ lines of ethylene at 915.5 cm$^{-1}$ and 922.5 cm$^{-1}$ (among several weaker lines) were modeled.  In the 620 cm$^{-1}$ - 640 cm$^{-1}$ region, the propyne $\nu_{9}$ band at 633 cm$^{-1}$ and diacetylene $\nu_{8}$ band at 628 cm$^{-1}$  were modeled.  
	
	The \textit{a priori} volume mixing ratios (VMRs) used were constant over our range of sensitive altitudes above 100 km and set to values comparable to those reported in previous literature including \cite{coustenis:2007}, \cite{vinatier:2007}, \cite{coustenis:2010}, and \cite{vinatier:2015}, and represent 'rough guesses' for the abundance of each molecule in both time spans.  Though some molecules have been shown to have vertical gradient that changes with time, we chose to use a constant profile above saturation that is the same for both time spans as to not influence the results of our comparisons across latitude and time.  The \textit{a priori} value for propene is comparable to predictions from \cite{hebrard:2013}, \cite{nixon:propene}, \cite{li:2015}, and \cite{dobrijevic:2016}.

	\section{Results}
\begin{figure}[h]
		\centering
		\includegraphics[height = \textheight]{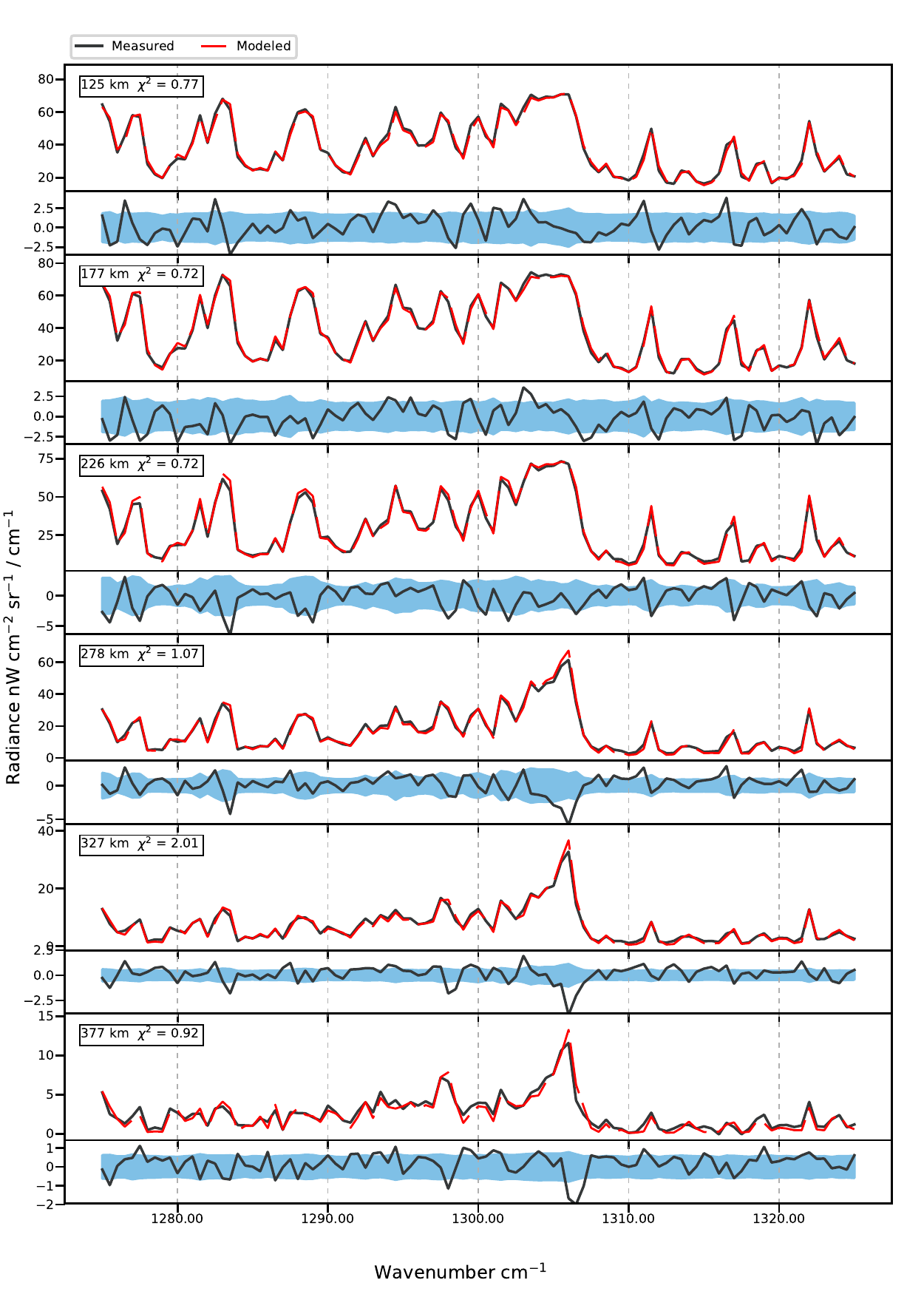}		
		\caption{Spectra used in the 2004-2009 equatorial temperature retrieval of the methane $\nu_{4}$ band.  The altitude label corresponds to the peak contribution altitude, see Fig. \ref{fig:tempcf}}
		\label{fig:tempfit}
	\end{figure}

\begin{figure}[h]
		\centering
		\includegraphics[width=\columnwidth]{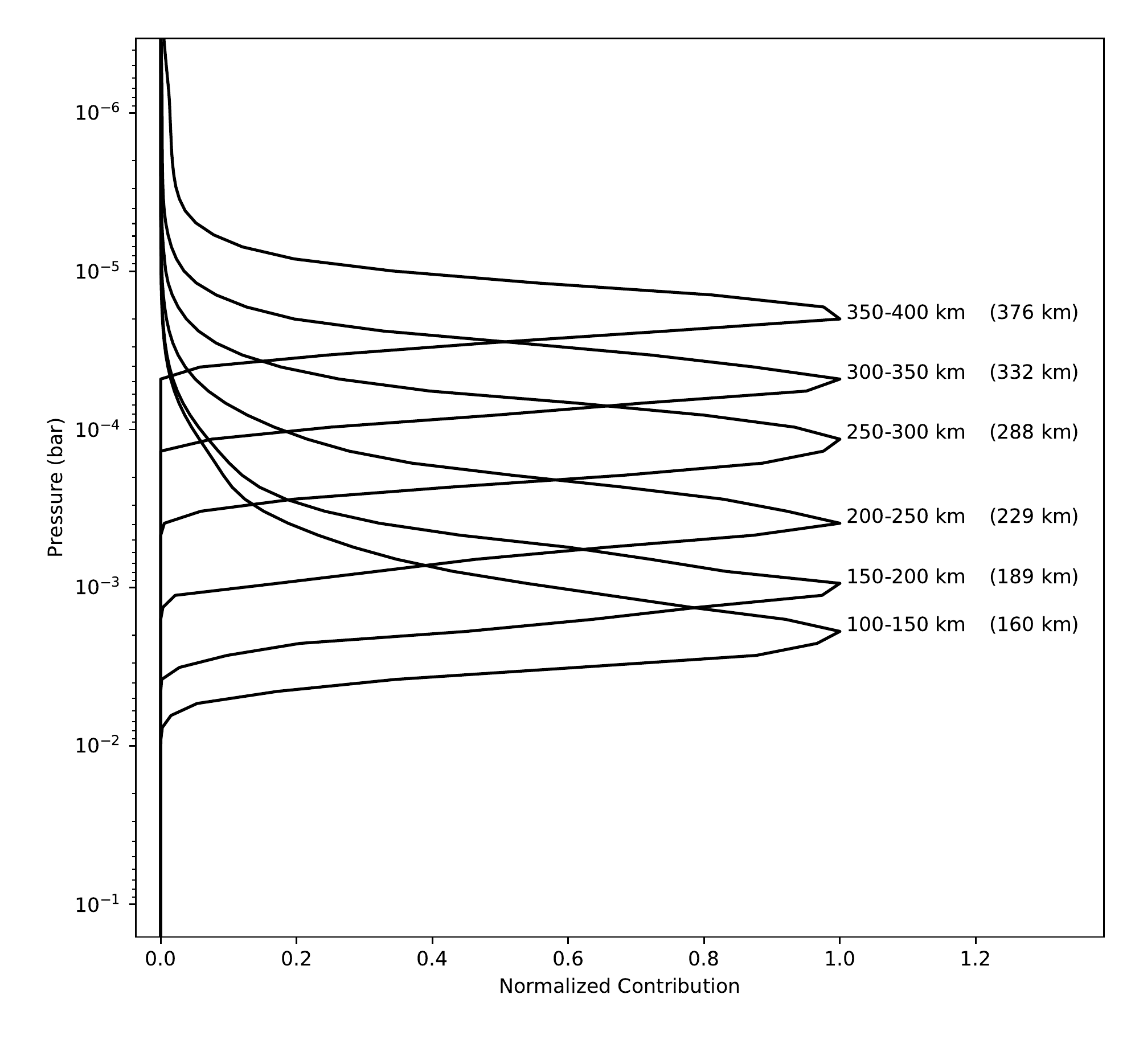}		
		\caption{Normalized temperature contribution functions for each altitude bin, taken at 1311 cm$^{-1}$.  Altitude bins are listed immediately to the right of each contribution function.  The altitude where the contribution function peaks is listed in parentheses.  In most cases, altitudes of peak contribution are not the center altitude for each bin.}
		\label{fig:tempcf}
	\end{figure}

\begin{figure}[h]
	\includegraphics[width=\columnwidth]{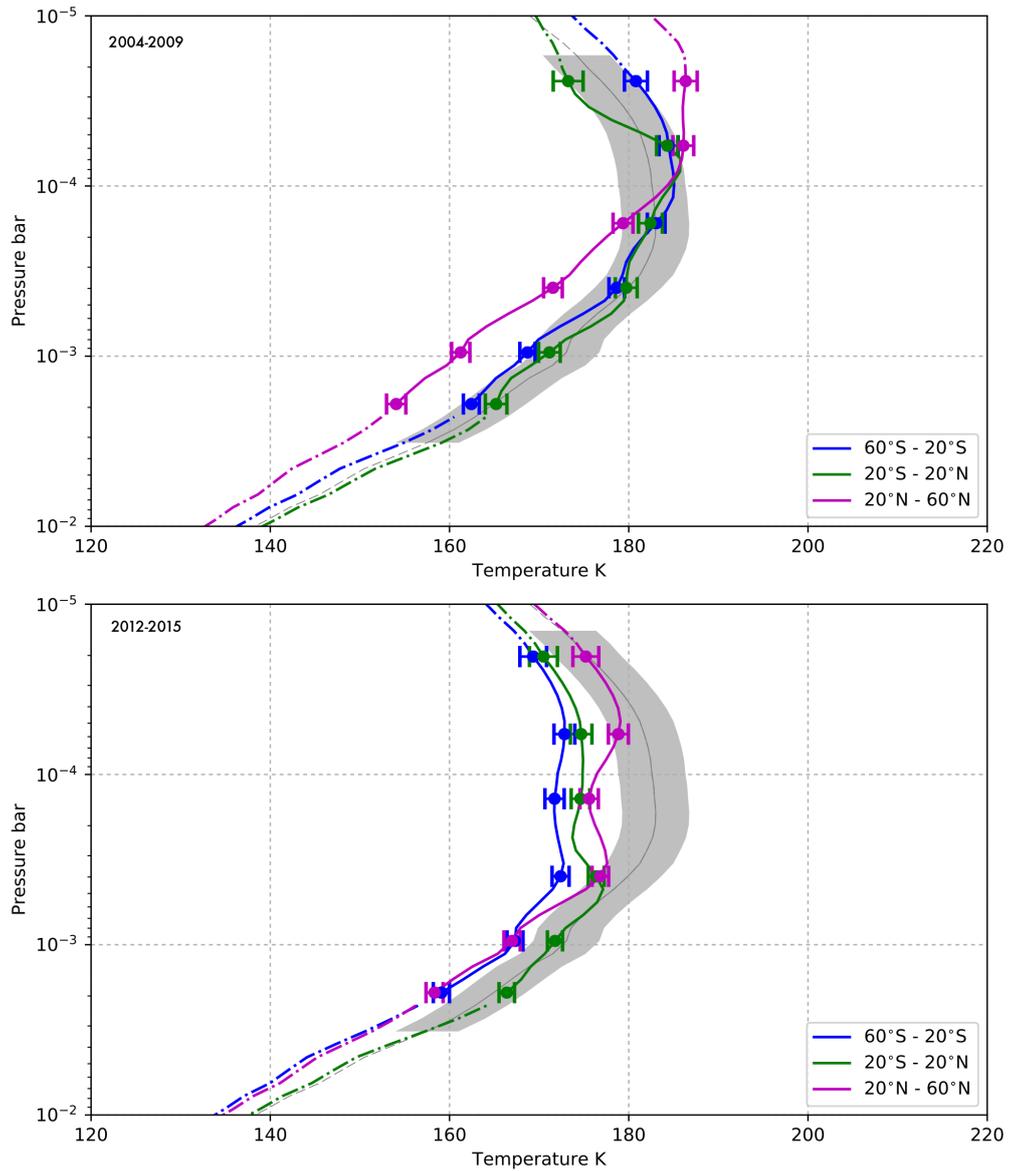}
	\caption{Comparison of retrieved temperatures in the 2012-2015 time span.  1-$\sigma$ error bars are given at altitudes where the contribution function peaks in each altitude bin.  The solid gray line is the a priori profile with error envelope.  Dot dashed lines are the retrieved profile where there are no spectra.}
	\label{fig:temps}		
\end{figure}

\begin{figure}[h]
	\centering
	\includegraphics[width=\columnwidth]{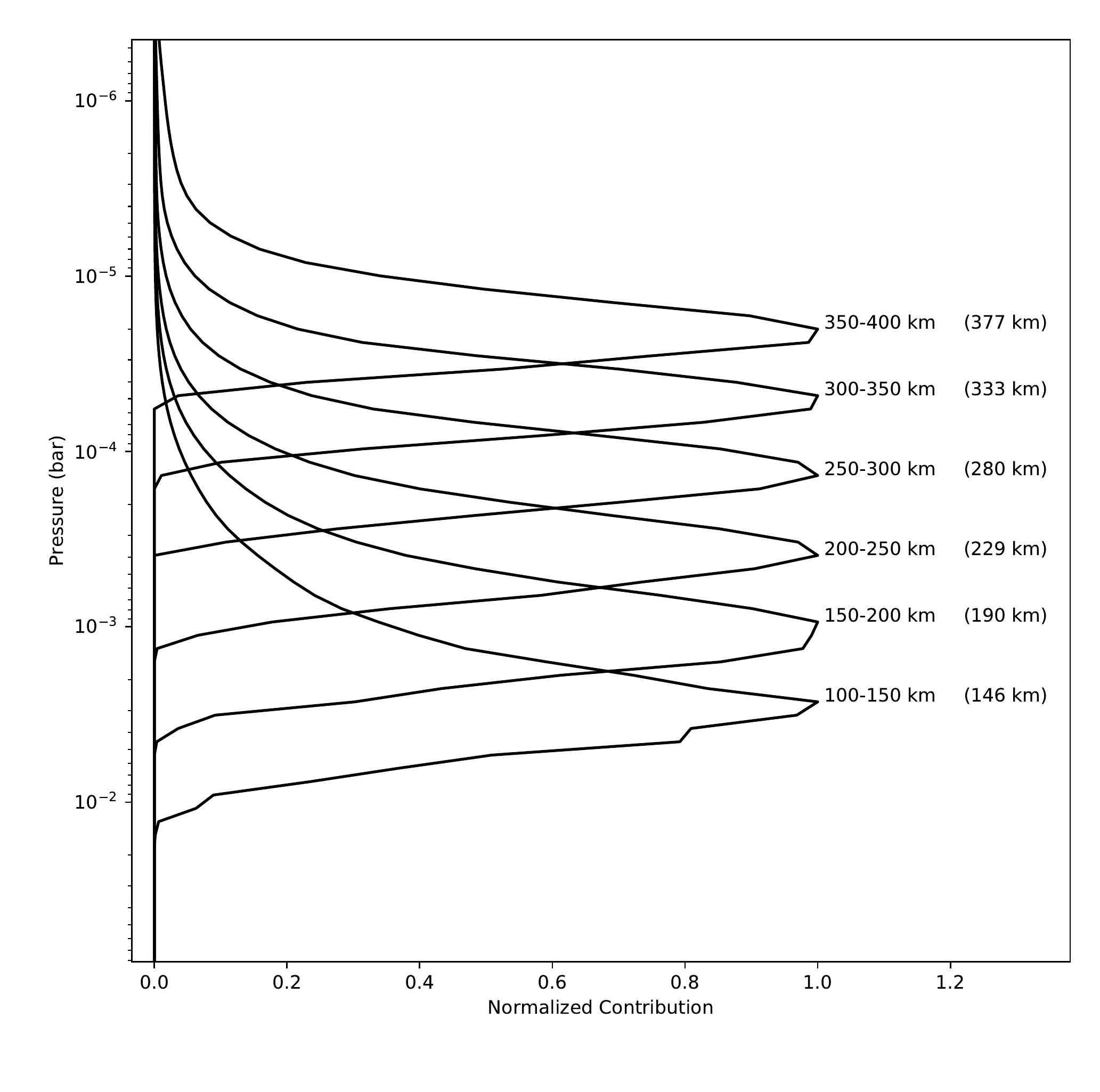}		
	\caption{Normalized propene contribution functions for each altitude bin, taken at 912.5 cm$^{-1}$.  Altitude bins are listed immediately to the right of each contribution function.  The altitude where the contribution function peaks is listed in parantheses.  In most cases, altitudes of peak contribution are not the center altitude for each bin.}
	\label{fig:propenecf}
\end{figure}

\begin{figure}[h]
		\centering
		\includegraphics[height=\textheight]{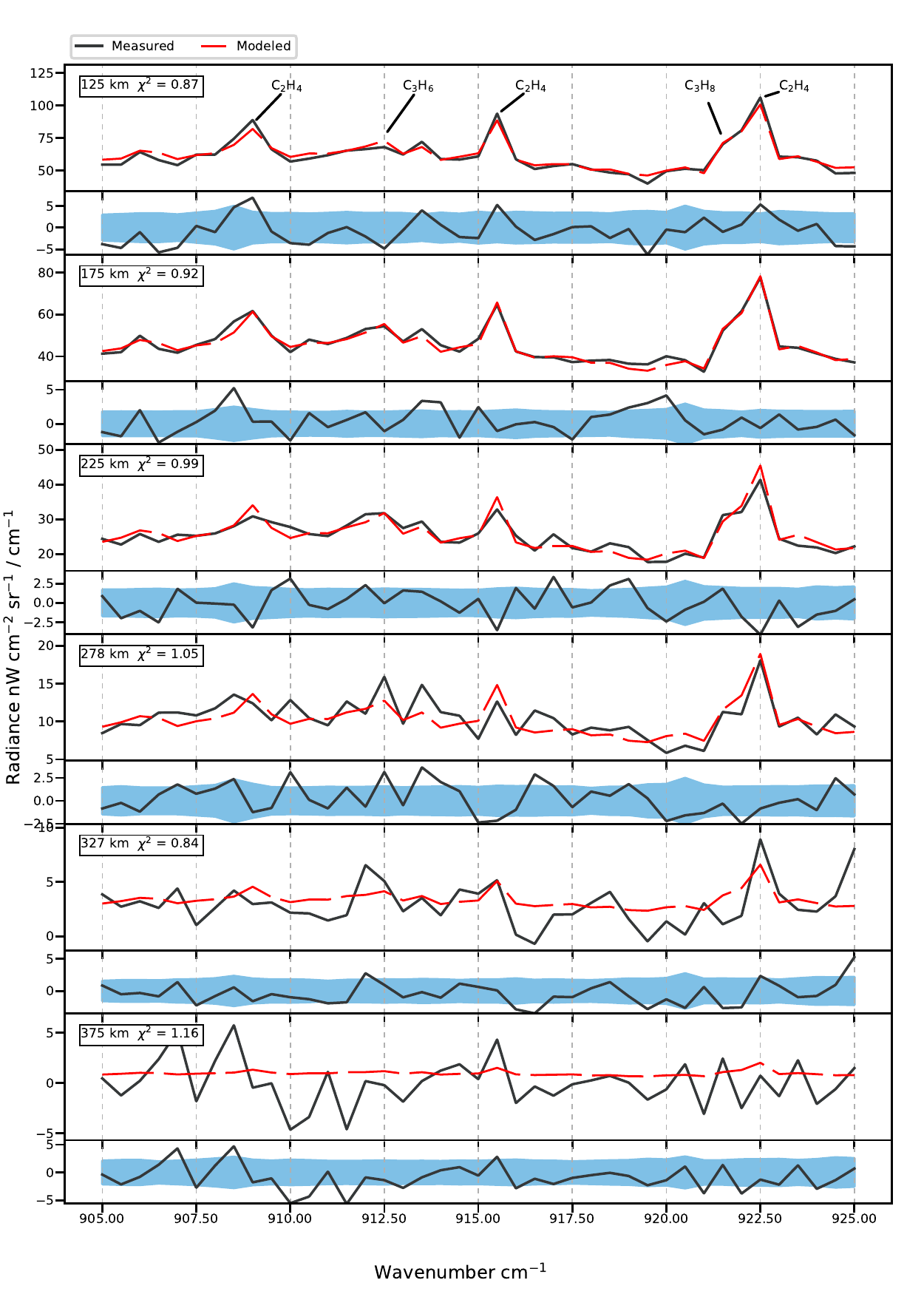}		
		\caption{Spectral fits for propene and propane for the 2004-2009 time span in the equatorial bin.}
		\label{fig:fp3spec}
	\end{figure}

\begin{figure}[h]
	\centering
	\includegraphics[height=\textheight]{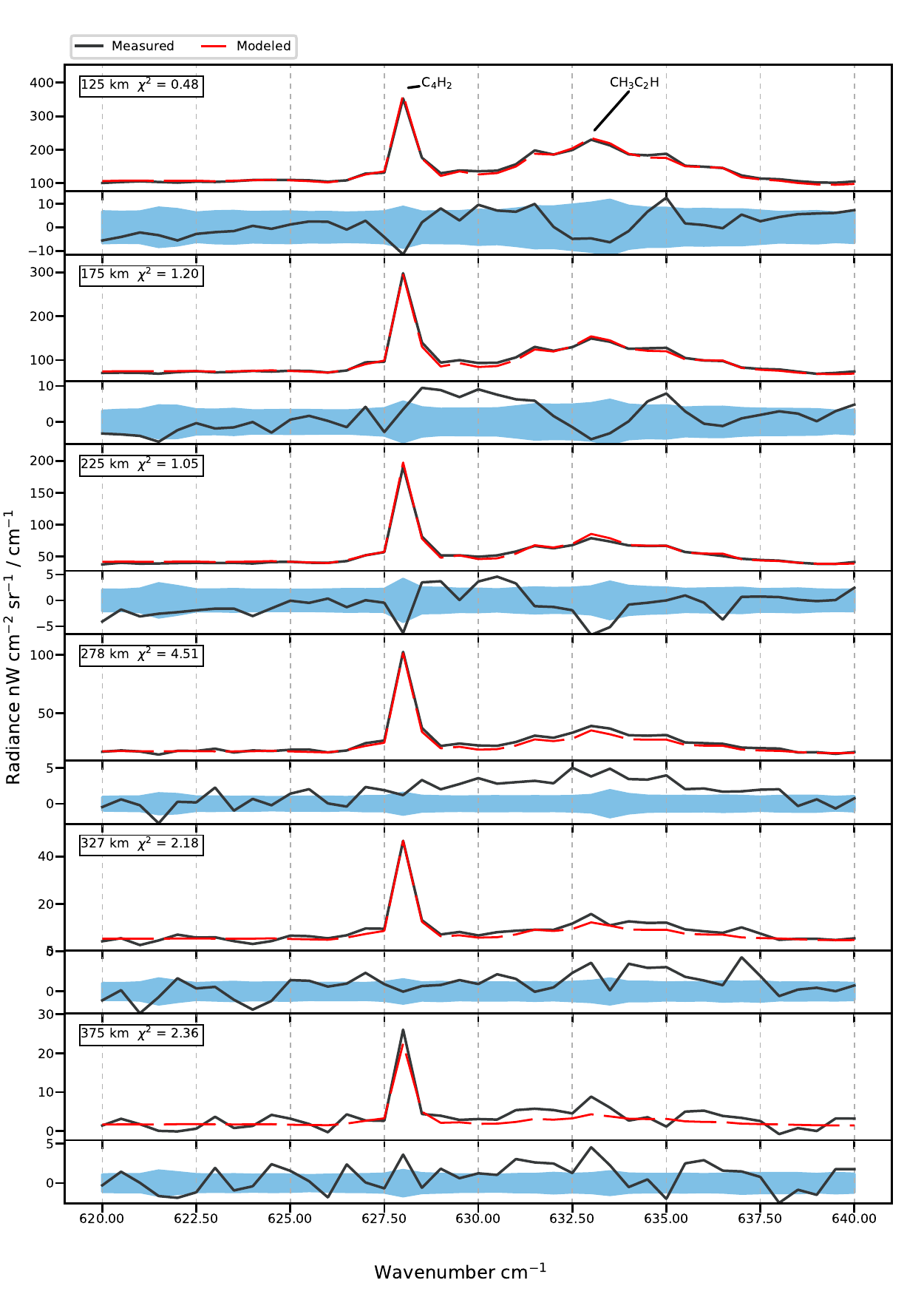}		
	\caption{Spectral fits used in the 2004-2009 south propyne retrieval.}
	\label{fig:propynespec}
	\end{figure}

\begin{figure}[p]
		\includegraphics[width=\columnwidth]{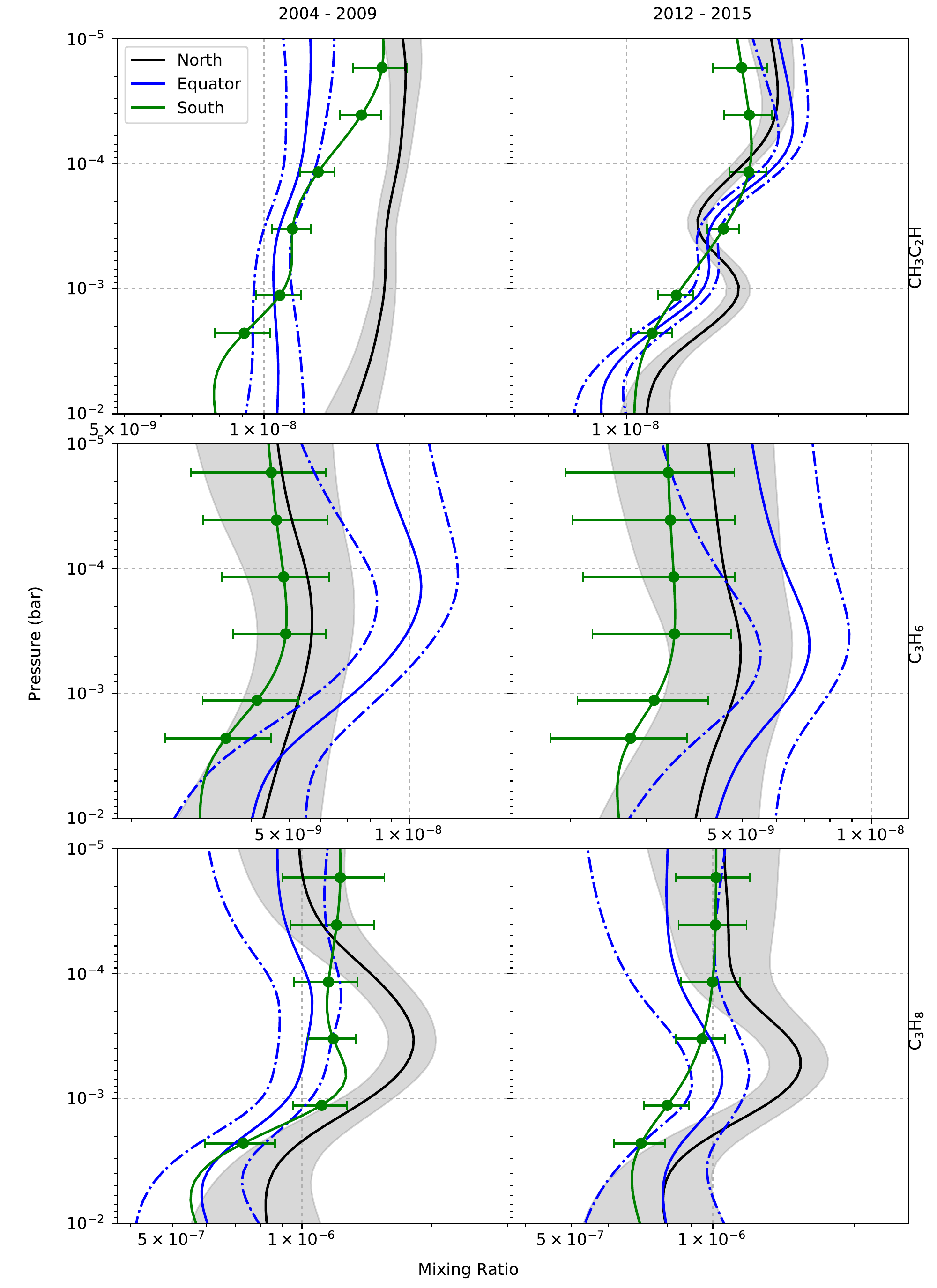}		
		\caption{Vertical profiles C$_{3}$H$_{4} $(top), C$_{3}$H$_{6}$ (middle), C$_{3}$H$_{8}$ (bottom). Black is the northern bin, blue is the equatorial bin, and green is the southern bin.  Volume mixing ratios are given as solid lines.  1-$\sigma$ errors are given as colored regions, dot-dashed lines, or horizontal error bars at altitudes of peak contribution.}
		\label{fig:timecompare}
	\end{figure}

\begin{figure}[p]
		\centering
		\includegraphics[height= \textheight]{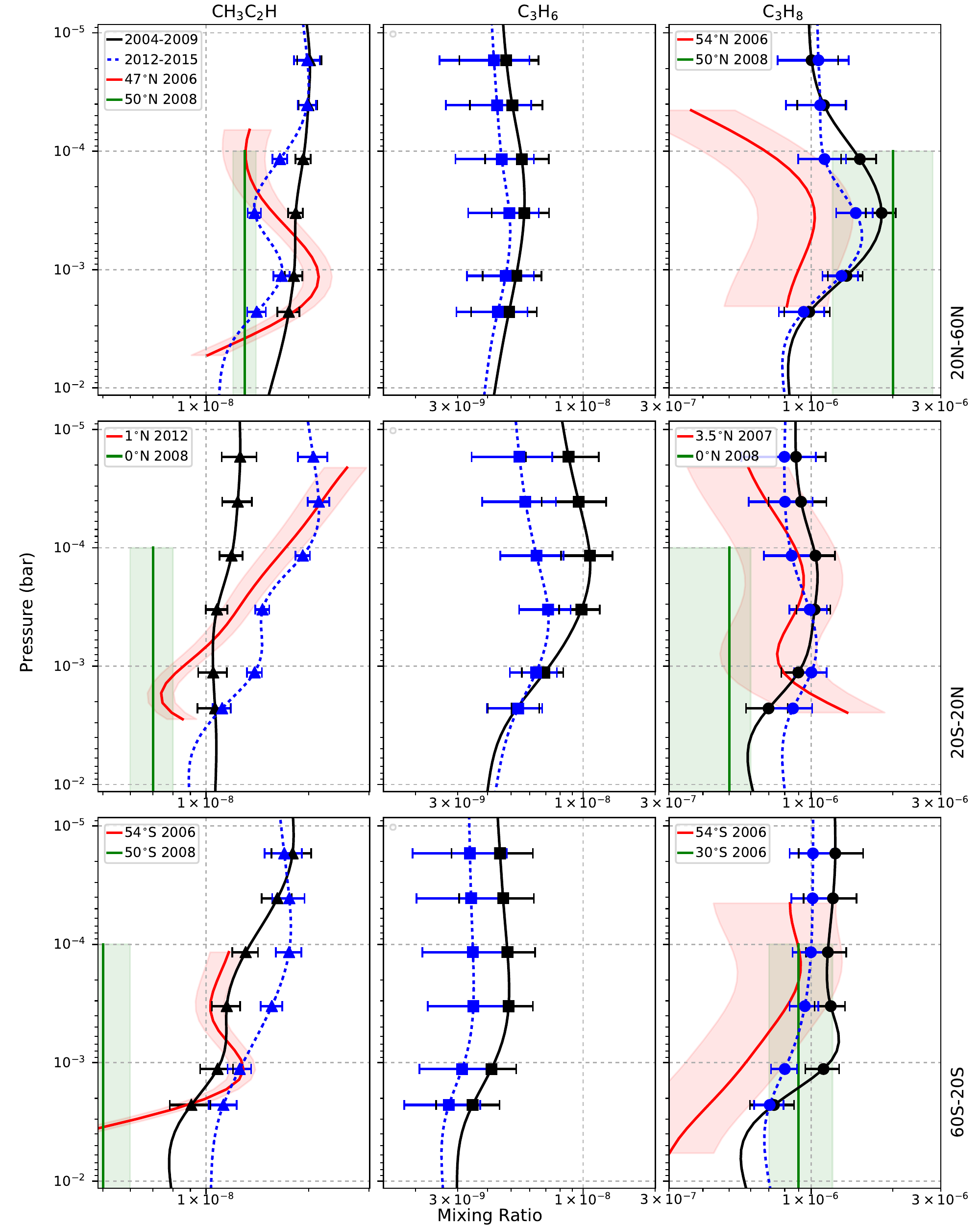}	
		\caption{Vertical profiles of our retrieved gases compared with profiles retrieved in \cite{vinatier:2015} and \cite{bampasidis:2012}. Latitude ranges are listed to the right.  Volume mixing ratios for the 2004-2009 time span are solid lines and volume mixing ratios for the 2012-2015 time span are dotted lines.  1-$\sigma$ errors are given as error bars at peak contribution altitudes.  Profiles from \cite{vinatier:2015} are shown in red with latitude and year of observation labeled and profiles from \cite{bampasidis:2012} are shown in green with latitude and year of observation labeled.}
		\label{fig:latcompare}
	\end{figure}

	Example spectra from the altitude bins used in the pre-equinox southern temperature retrieval are shown in Fig \ref{fig:tempfit}.  Errors on the radiance were initially calculated as the standard deviation from the mean of all spectra included in the average, but were expanded to account for systematic uncertainties and prevent over-constraining  the model.  Normalized contribution functions - also called the inversion kernel -  are the rate of change of radiance with respect to abundance, normalized to one at the maximum value. Altitudes with higher values 'contribute' more to the calculated spectrum, and thus are the altitudes where the data give the most information - see Fig. \ref{fig:tempcf}.  Retrieved temperature profiles for the pre-equinox time span and the post-equinox time span are shown in Fig. \ref{fig:temps}.  We see variation in the stratospheric temperature and the shape and altitude of the stratopause as described in \cite{flasar:2009}.  While the stratopause increases in altitude towards the north winter pole, we do not see the full extent of the stratopause, since our data is only sensitive to an altitude of 0.02 mbar.
	Contribution functions similar to Fig. \ref{fig:tempcf} are shown in Fig. \ref{fig:propenecf} for the propene retrieval.
	Figs. \ref{fig:fp3spec} and \ref{fig:propynespec} show spectral fits for the pre-equinox southern bin, similar to those shown for temperature.
	The retrieved profiles of C$_{3}$H$_{4}$, C$_{3}$H$_{6}$, C$_{3}$H$_{8}$ are given in Fig. \ref{fig:timecompare} and Fig. \ref{fig:latcompare}, and are compared to profiles reported in \cite{bampasidis:2012} and \cite{vinatier:2015} in Fig. \ref{fig:latcompare}.
	
	The C$_{3}$H$_{8}$ profile shows a peak stratospheric abundance at 0.5 mbar in the north during late winter, with the profile flattening out to a more vertically constant profile in the equatorial and southern bins.  The post-equinox profiles show the peak altitude decreasing slightly in the north, and a similar peak at 0.7 mbar forming around the equator during early northern spring.
	 	
	C$_{3}$H$_{6}$ does not follow the same trend as other C$_{3}$ hydrocarbons, the maximum abundance of the gas in both time seasons is above 1 mbar near the equator.  This differs from the general trend of the other C3 hydrocarbons and trace gases, which tend to increase above the mid to high winter latitudes. 
	
	CH$_{3}$C$_{2}$H shows a nearly constant abundance within error bars in the northern pre-equinox bin, becoming more variable and decreasing in abundance as the latitude moves away from the winter pole.  The CH$_{3}$C$_{2}$H abundance profiles for the post-equinox time span are comparable across latitudes.  There appears to be some small but systematic error of the CH$_{3}$C$_{2}$H spectral fit that may require re-evaluation of the methylacetlyene line list.  This misfit is also seen in Fig. 10 of \cite{vinatier:2007} and Fig. 12 of \cite{coustenis:2007}.
	
	\section{Discussion}
	\subsection{Propane}
	
	In the pre-equinox period, equatorial and southern propane increases with altitude from 5$\times$10$^{-7}$ at 10 mbar to 1$\times$10$^{-6}$ at 0.01 mbar.  In the north (during winter at the time), propane achieves a maximum abundance of 1$\times$10$^{-6}$ at 0.3 mbar before it begins to decrease with altitude, returning to an abundance comparable to the southern and equatorial regions.  In the post-equinox period, the propane distribution remains largely unchanged within error bars.  The south shows the abundance around 1 mbar decrease from 1$\times$10$^{-6}$ to 8$\times$10$^{-7}$, just outside of the error bars on the retrieval.
	
	We are able to compare to \cite{vinatier:2007}, who measure abundance from the first flyby of Titan by Cassini, Tb on 13 December 2004, at 15 $^{\circ}$S.  They show propane increasing monotonically from 4$\times$10$^{-7}$  at 4 mbar to 1$\times$10$^{-6}$  at 0.01 mbar, which has a slightly greater slope than our pre-equinox values.  In comparison, \cite{bampasidis:2012} probe the lower stratosphere using nadir observations between 2006 and 2012 and show propane generally at a southern and equatorial abundance of 5$\times$10$^{-7}$  between 2006 and 2009, in agreement with our results.  At 50$^{\circ}$N, they retrieve an abundance near 2$\times$10$^{-6}$ , which is slightly higher than our values at the lowest altitudes, but agree with our limb sounding around 3 mbar.  \cite{vinatier:2015} perform another analysis of chemical abundance during the northern spring, between 2009 and 2013.  At 46$^{\circ}$N in 2012, they show propane  increase from 4$\times$10$^{-7}$  at 5 mbar to 1$\times$10$^{-6}$ at 0.03 mbar.  Because their analysis was done on a single observation, the error bars presented are large, so it is hard to compare variations in the vertical profile with our results, however they do seem compatible.
	
	\subsection{Propyne}
	
	Propyne shows stronger seasonal variation than propane.  In the pre-equinox period, our results show propyne having a weak vertical gradient from 8$\times$10$^{-9}$  at 8 mbar to just under 2$\times$10$^{-8}$ at 0.01 mbar in the south.  The vertical gradient decreases near the equator, and becomes nearly vertical in the north, indicative of the effect of downward advection within the winter polar vortex on the abundance profiles of trace gases. In the post-equinox period, the propyne profile behaves similar across all latitudes, increasing from 1$\times$10$^{-8}$ at 8 mbar to 2$\times$10$^{-8}$ at 0.01 mbar.  \cite{bampasidis:2012} report similar values to ours for the 2006-2009 period for 50$^{\circ}$N, 0$^{\circ}$N, and 50$^{\circ}$S.  \cite{vinatier:2007} show a steeper vertical profile at 15$^{\circ}$S than our results indicate.  Their modeled profile begins at a slightly lower abundance at 1 mbar and is in agreement with our results at higher altitudes.  This could be explained by our larger dataset including observations later in the season and at slightly higher latitudes, where we would expect to see more propyne at lower altitudes.

	\subsection{Propene}	
	Photochemical models have included C$_{3}$H$_{6}$ beginning after the Voyager flyby \citep{yung:1984}.  Since the start of the Cassini mission, several new models have also incorporated the molecule \citep{wilson:2004, hebrard:2013, kras:2014, li:2015, loison:2015, dobrijevic:2016}.
	
	\cite{yung:1984} propose the main source of propene in Titan's stratosphere 
	(<300 km) is from 
	
	\begin{equation}
	\label{}
		C_{2}H_{3} + CH_{3} + M \rightarrow C_{3}H_{6} + M
	\end{equation}
	
	\begin{equation}
	\label{}
	C_{2}H_{3} + CH_{3} + M \rightarrow C_{3}H_{6} + M
	\end{equation}
	\cite{li:2015} follow with additional production pathways 
	
	\begin{equation}
	CH + C_{2}H_{6} \rightarrow C_{3}H_{6} + H
	\end{equation}
	dominating the region between 600 km and 1000 km, where CH molecules are plentiful and
	
	\begin{equation}
	C_{3}H_{8} + h\nu \rightarrow C_{3}H_{6} + H_{2}
	\end{equation}
	dominating the region between 300 km and 600 km, where the pressure is not great enough for a termolecular reaction and CH is scarce.  Alternatively, \cite{hebrard:2013} include 
	
	\begin{equation}
	H + C_{3}H_{5} \rightarrow C_{3}H_{6}
	\end{equation}
	as the dominating production reactions at mid to high altitudes.
	
	The first detection of C$_{3}$H$_{6}$ in Titan's atmosphere was made with the INMS instrument by \cite{magee:2009}.  However, because mass spectrometers can generally not differentiate between isomers, it is unknown whether this detection is of propene or cyclopropane.  The first definitive detection of propene was made by \cite{nixon:propene} using an average of CIRS spectra between 30$^{\circ}$S and 10$^{\circ}$N from 1 July 2004 through 1 July 2010, similar to our 20$^{\circ}$S-20$^{\circ}$N pre-equinox bin.  However, a spectral line list for propene did not exist at the time, so the gas could not be included in their radiative transfer calculations.  Instead, estimates of abundance were made by comparing the relative intensities of the propane and propene lines, and they claimed a 3-$\sigma$ abundance of (2.0-4.6)$\times$10$^{-9}$ between 100 and 200 km. 
	
\begin{figure}[h]
		\centering	
		\includegraphics[width=\columnwidth]{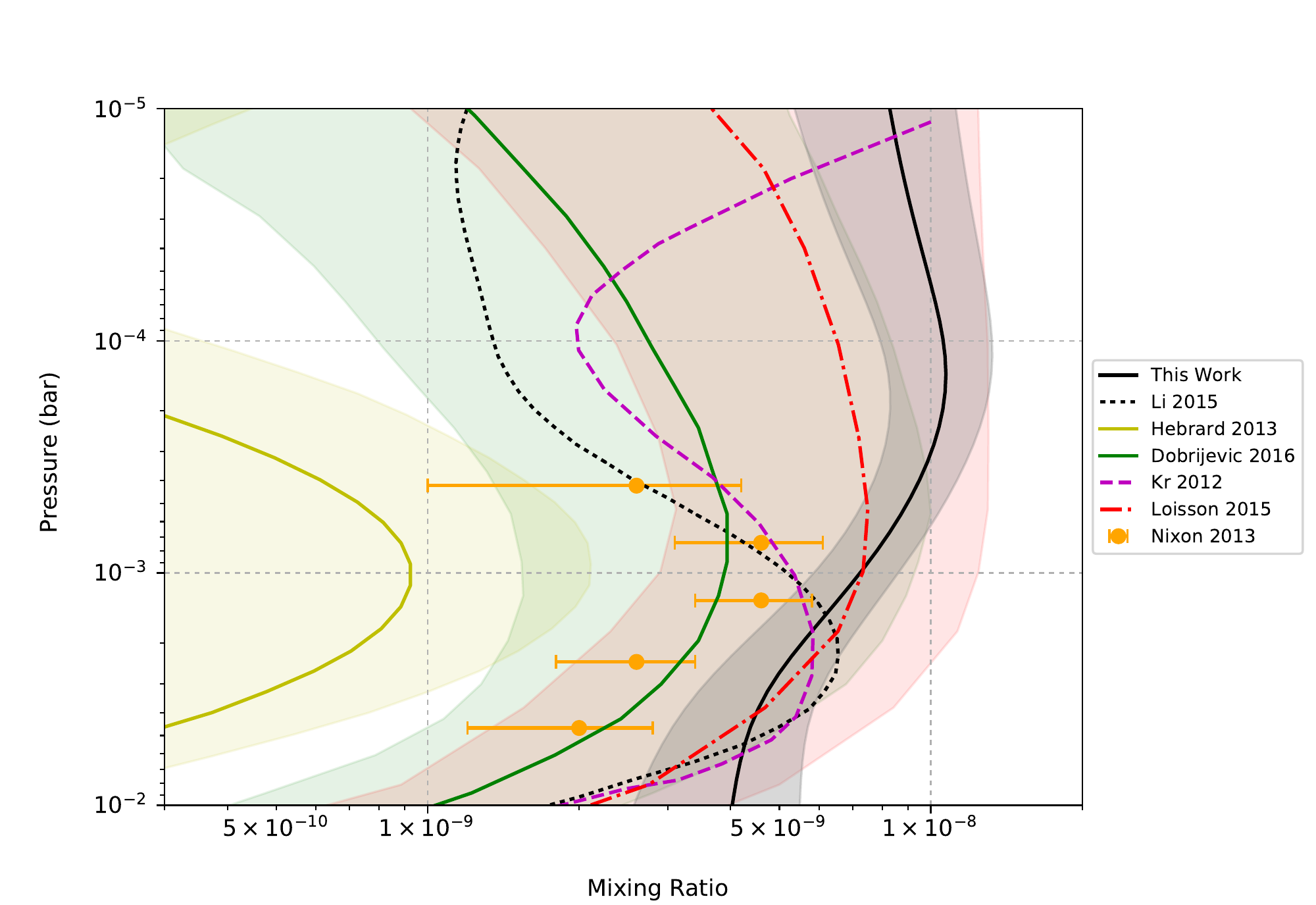}
	
		\caption{Our 2004-2009 20$^{\circ}$S - 20$^{\circ}$N profile compared with published predictions for propene's abundance.  Errors on Hebrard 2013 correspond to 75th percentile, Loison 2015 correspond to 90th percentile, Dobrijevic 2016 correspond to 90the percentile.  Nixon 2013 values are inferred abundances and uncertainties described in \cite{nixon:propene}.  Of the models compared, only Loison 2015 has a similar shape and abundance to our measured values.  Above 10$^{-4}$ bar, the signal from propene drops to near the NESR, so a the abundances presented here are most reliable below this altitude.}
		\label{fig:licompare}
	\end{figure}

	In Fig. \ref{fig:licompare}, we compare the predicted propene profiles to the abundance we determined for the pre-quinox time span, centered on the equator.  Of the models compared, \cite{loison:2015} shows the best agreement.  The abundance below 6 mbar is within the error bars of our measured values.  Above that, our measurements are within the 90th percentile of modeled profiles (90th percentile being the region which encloses 90\% of modeled profiles.)  In all other cases, the predictions have a peak abundance at or below 1 mbar, ranging from 1 to 6 ppbv.  Our profiles display a peak abundance of 10 ppbv at 0.1 mbar, higher in abundance and altitude than all of the compared profiles.  The profiles from \cite{li:2015} and \cite{kras:2014} show good agreement in abundance below 1 mbar, but also display an inversion above 0.2 mbar that we do not see in our measurements.  We also retrieve about twice as much propene compared to the values inferred from relative line strengths in \cite{nixon:propene}.
	
	A modified version of the \cite{loison:2015} model was used as the \cite{dobrijevic:2016} model.  Since the two models have different predicted profiles for propene, we can look at the differences between the two models in attempt to isolate the factors that caused the abundance of propene to change.  \cite{dobrijevic:2016} list the changes from \cite{loison:2015} as: limiting the modeled hydrocarbons at C$_{4}$ species, excluding high mass nitriles, reducing the number of isomers included, and not considering reactions with 'very small fluxes'.  The authors checked the effects of these changes on species included in the model, and while no major change was reported for significant molecules, we do see that the changes applied in \cite{dobrijevic:2016} decrease the predicted abundance of propene and worsen the agreement between the \cite{loison:2015} and our measurements.  Due to the large number of reactions included in both models, we are unable to say which of these changes has the greatest effect on the modeled abundance of propene.
	
	\subsection{Allene}
	The isomer of propyne, CH$_{2}$CCH$_{2}$ or allene, is also theorized to be produced in Titan's atmosphere \citep{yung:1984, li:2015}.  Production pathways for allene (and propyne) include 
	
	\begin{equation}
	CH + C_{2}H_{4} \rightarrow C_{3}H_{4}
	\end{equation}
	above 600 km, with
	
	\begin{equation}
	H + C_{3}H_{5} \rightarrow C_{3}H_{4} + H_{2}
	\end{equation}
	 and
	 
	 \begin{equation}
	CH_{3} + C_{3}H_{5} \rightarrow C_{3}H_{4} + CH_{4}
	\end{equation}
	dominating throughout the rest of the atmosphere.
	
	The only potential detection of allene on Titan was made in \cite{roe:2011}, however since the line list used in the authors' analysis was in significant disagreement with the potential observed allene lines this remains only a tentative detection.  Other analyses of Titan's atmosphere to search for allene have resulted in many upper limits.  \cite{coustenis:2003} claimed a 3-$\sigma$ upper limit of 2 ppbv, derived from disk averaged spectra from the Infrared Space Observatory (ISO).   \cite{nixon:allene} estimate a 3-$\sigma$ upper limit of 0.3 ppbv at 25$^{\circ}$ N at 107 km, and 1.6 ppbv at 76$^{\circ}$ N, 224 km, using a method described in their paper.  As discussed in Section 2.2.1, the line data used in these previous works is inaccurate.  We update these upper limits using a corrected line list produced by us and included in our radiative transfer calculations.
	
\begin{figure}[p]
	\includegraphics[height=\textheight]{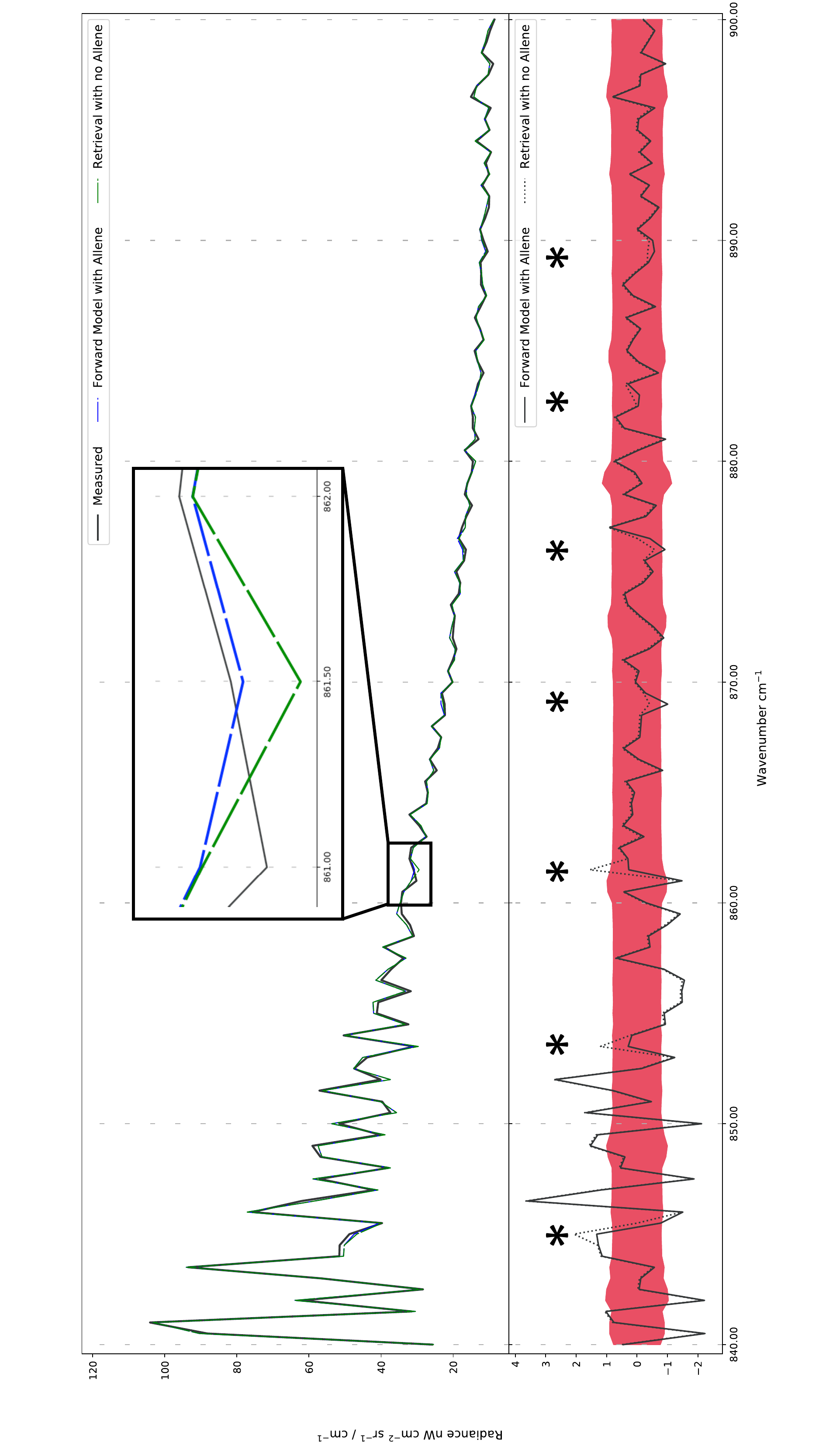}
	\caption{Comparison of modeled spectrum without allene (green) and forward modeled spectrum including allene (blue), against the observed spectrum (black).  Differences in the spectra are most easily noted in the residuals, where contributions from allene are seen as excess emission in the dashed spectrum, noted by asterisks.}
	\label{fig:allenespec}
	\end{figure}

	The spectrum used in the model is an average of nadir observations between 30$^{\circ}$S and 30$^{\circ}$N, from 2004-2015.  Nadir observations were used to take advantage of the lower noise compared to limb viewing, however the abundance of allene is likely to be very low at the altitudes that nadir observations are sensitive to. Fig. \ref{fig:allenespec} shows the best fit spectrum overlayed on the observed spectrum. The contributions of allene are visible as spikes in the dotted-line residual.
	
\begin{figure}[p]
	\includegraphics[width=\columnwidth]{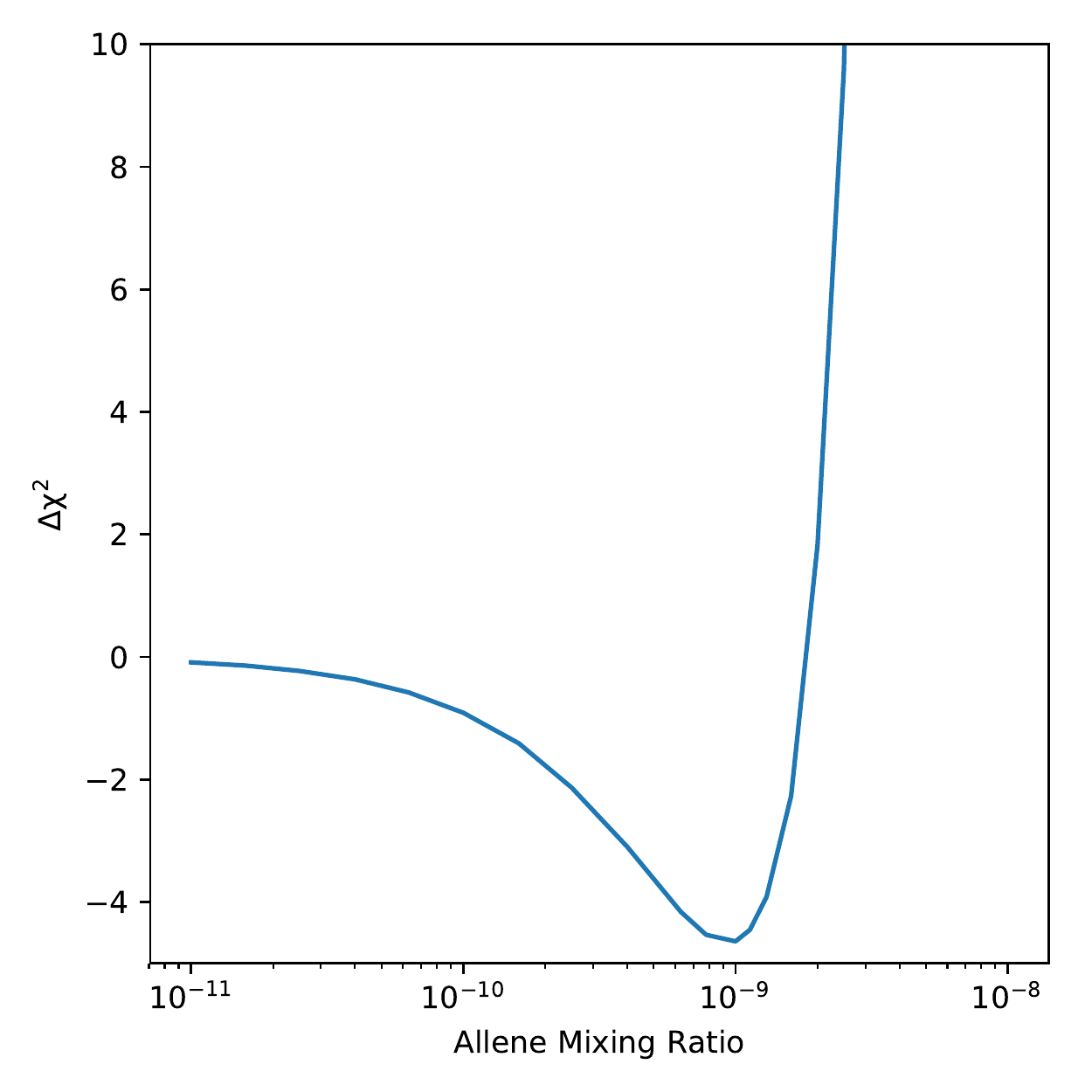}
	\caption{Plot of the $\Delta \chi^{2}$ against allene abundance.  $\Delta \chi^{2}$ is calculated by subtracting the  $\chi^{2}$ calculated from the forward model with allene included in the atmosphere from the $\chi^{2}$ value calculated from retrieved spectrum that does not include allene.  Along these lines, a negative $\Delta \chi^{2}$ indicates the model-fit has improved, whereas a positive value indicates the fit is worsened.  The $\Delta \chi^{2}$ value achieves a value of 9 near 2.5 $\times$10$^{-9}$, leading us to an estimated 3-$\sigma$ upper limit of 2.5 ppbv.}
	\label{fig:allenechi}
	\end{figure}

	We perform a $\Delta \chi ^{2}$ analysis similar to that described in \cite{teanby:2007} and \cite{nixon:allene} and estimate an updated 3-$\sigma$ upper limit of 2.5 ppbv at 150 km within 30$^{\circ}$ north and south of the equator.  A retrieval of ethane, propane, and an aerosol haze was performed between 830 and 880 cm$^{-1}$, with allene line data not included.  The $\chi^{2}$ value for this retrieval was  considered $\chi_{0} ^{2}$.  The best-fit abundances of ethane and propane were then fixed and varying amounts of allene were added to the atmosphere model. A forward model was run at each allene abundance to calculate a modified chi-squared value ($\chi_{m}$), the $\Delta \chi ^{2}$ value was calculated to be $\chi_{m} ^{2}$ - $\chi_{0} ^{2}$.  We plot the values of $\Delta \chi ^{2}$ as a function of added allene in Fig.  \ref{fig:allenechi}.  Where the $\Delta \chi ^{2}$ achieves a value of 9, we claim an upper limit with a confidence of $(\Delta \chi ^{2})^{1/2}$ = 3-$\sigma$. The results are shown in Fig. \ref{fig:allenespec} and Fig. \ref{fig:allenechi}. A major challenge to modeling this allene band is the overlap of the band with the band of ethane and the band of propane.  The ethane band is very bright, as seen in Fig. \ref{fig:allenespec}.  Therefore, relatively small discrepancies in the modeling and line list of ethane propagate heavily through to measuring allene.  The propane band also present in the region we modeled is very dim and broad, further increasing the difficulty of measuring allene in this region, since it contributes mostly to the continuum in this small spectral window.
	
	\section{Conclusion}
	In this work, we have determined the first abundance profiles of propene in Titan's atmosphere, enabling us to compare the mixing ratio and spatial distribution of this gas with other trace gases, as well as compare to predictions from existing photochemical models.  We've shown:
	
	\begin{enumerate}
		\item propene is present in Titan's atmosphere at a mixing ratio between 4 and 10 ppbv in the stratosphere
		\item the abundance of propene near the equator is consistent with predictions from the \cite{loison:2015} model, but other predictions show a local inversion centered around 0.1 mbar, and a generally lower abundance
		\item contrary to other trace gases, propene does not show a poleward enhancement in the winter hemisphere equatorward of 60$^{\circ}$.  Instead, propene shows an increased abundance above the equator relative to either pole.
	\end{enumerate}

	The results of our analysis will be useful in refining models of Titan's atmospheric chemistry and dynamics as discrepancies between observations and predictions show that current photochemical models do not accurately predict the production and destruction rates of the molecule.  The unique spatial trend exhibited by propene will make it a useful constraint in global circulation and transport models, as it may be a good tracer for horizontal transport.  Additionally, the polymerization of propene and other $\pi$-bond molecules may lead to the formation of Titan's haze\citep{teanby:haze, trainer:2013}.

	We also provide a new 3-$\sigma$ upper limit for allene within 30$^\circ$ of the equator of 2.5$\times10^{-9}$.  Because this value is calculated using corrected line data, it should be considered more reliable  than values from previous calculations.
	
	\section{Acknowledgments}
	N.A.L., C.A.N., R.K.A., and F.M.F. were supported by the NASA Cassini Project for the research work reported in this paper.  C. A. N. also acknowledges support from the NASA Astrobiology Institute.  Part of the research was carried out at the Jet Propulsion Laboratory, California Institute of Technology, under a contract with the National Aeronautics and Space Administration.

	\clearpage
	\section{References}
	\bibliography{c3}
	\bibliographystyle{elsarticle-harv}
	
\end{document}